\documentclass[twoside,twocolumn,tightenlines, showpacs,aps,prb]{revtex4}
\usepackage{graphicx}
\usepackage{mathrsfs}
\usepackage{cancel}
\newcommand{\be}{\begin{equation}}
\newcommand{\ee}{\end{equation}}

\begin{document}

\title{Conductance Peak Density in Nanowires}

\author{T. Ver\c{c}osa$^{1,3}$, Yong-Joo Doh$^2$, J. G. G. S. Ramos$^3$, A. L. R. Barbosa$^1$}

\affiliation{$^1$ Departamento de F\'{\i}sica,Universidade Federal Rural de Pernambuco, 52171-900 Recife, Pernambuco, Brazil\\$^2$ Department of Physics and Photon Science, Gwangju Institute of Science and Technology (GIST), 123 Cheomdan-gwagiro, Gwangju 61005, Republic of Korea\\$^3$ Departamento de F\'{\i}sica, Universidade Federal da Para\'iba, 58297-000 Jo\~ao Pessoa, Para\'iba, Brazil }

\date{\today}

\begin{abstract}

We present a complete numerical calculation and an experimental data analysis of the universal conductance fluctuations in quasi-one-dimension nanowires. The conductance peak density model, introduced in nanodevice research on [Phys. Rev. Lett. 107, 176807 (2011)], is applied successfully to obtain the coherence length of InAs nanowire magnetoconductance and we prove its equivalence with correlation methods. We show the efficiency of the method and therefore a prominent alternative to obtain the phase-coherence length. The peak density model can be similarly applied to spintronic setups, graphene and topological isolator where phase-coherence length is a relevant experimental parameter.

\end{abstract}
\maketitle

\section {Introduction} 

Universal observable fluctuations are one of the most important fingerprints of chaos in quantum scattering processes [\onlinecite{Stockmann,Verbaarschot,Guhr,Mitchell,Lee,Mello,HoutenPRB,Houten,Beenakker,Alhassid,Akkermans,Kumar,Cheng}]. They began to be studied in compound nucleus scattering due to the emerging random fluctuations in response to energy variation in the cross-section [\onlinecite{Verbaarschot,Guhr,Mitchell}]. The universal fluctuation is usually characterized in terms of the correlation function $C(\delta Z) = \left\langle G(Z)G(\delta Z) \right\rangle -\left\langle G(Z) \right\rangle^2$ as a function of an arbitrary parameter $Z$. The correlation function measures the degree of coherence present in the otherwise fully chaotic system. The degree of coherence resides in the correlation width length $\Gamma_Z$, which defines the shape of correlation function. The cross-section correlation has been calculated by Ericson [\onlinecite{Ericson}] and renders a Lorentzian shape as a function of the energy variation, $C(\delta \epsilon) \propto [1+\left(\delta\epsilon/\Gamma_\epsilon\right)^2]^{-1}$. 

In nuclear physics, Brink and Stephen [\onlinecite{Brink}] proposed a simple model to access the correlation width length without the laborious experimental data obtention associated to the calculating of the correlation function. The model is founded on counting the number of maxima featured from the cross section as a function of the energy. Furthermore, they showed that resonance peak density (number of maxima per unit resonance energy) is given by $\rho_{\epsilon} =  \sqrt{3}/(\pi\Gamma_{\epsilon})$ in the limit of large number of reaction channels. The density peak method could access the width $\Gamma_{\epsilon}$ with more accurate and less experimental data than from the cross correlation function [\onlinecite{Dietz}] that requires a large ensemble.

The density peak  model was recently introduced in the nanodevice research in Refs.[\onlinecite{Ramos,Barbosa}] inspired by the formal analogy between conductance and compound-nucleus Ericson fluctuations. They found that the conductance peak density of nanodevices holds the same result of compound nucleus for energy variation. However, modern nanodevices as nanowires [\onlinecite{Lee,Houten,Akkermans,Doh,Alagha, Kim,Elm}], open quantum dots  [\onlinecite{Jacquod,Miller,Blumel,Jalabert,Efetov,Vallejos}] and graphene flakes  [\onlinecite{Kharitonov,Tikhonenko,Ojeda-Aristizabal,Ramos2,Lundeberg,Horsell}] enables measuring the conductance as a function of other external controllable parameters as perpendicular and parallel magnetic fields. 

Using the correlation function for perpendicular magnetic field variation [\onlinecite{Efetov}], square Lorentzian $C(\delta B_\perp) \propto [1+\left(\delta B_\perp/\Gamma_\perp\right)^2]^{-2}$, the Ref. [\onlinecite{Ramos}] has shown that the conductance peak density is given by 
\begin{eqnarray}
\rho_\perp = \frac{3}{\pi \sqrt{2}\Gamma_\perp} \approx\frac{0.68}{\Gamma_\perp},\label{rhopp}
\end{eqnarray}
while the Ref.[\onlinecite{Barbosa2}] found
\begin{eqnarray}
\rho_{||} =  \frac{\sqrt{3}}{\pi \Gamma_{||}} \approx\frac{0.55}{\Gamma_{||}},\label{rhopr}
\end{eqnarray}
for parallel magnetic field variation. The Eqs.(\ref{rhopp}) and (\ref{rhopr}) give rise to efficient alternative methodology in nanodevices to obtain the correlation width length instead of correlation function which requires a large number of realizations to be calculated. 

Motivated by findings of Refs.[\onlinecite{Ramos,Barbosa}], Dietz {\it et al.}  [\onlinecite{Dietz}] made the first test of peak density method experimentally with unprecedented accuracy, analyzing the cross-section fluctuations in open microwave billiard and quantum graphs. Furthermore, the density of maxima was introduced as a novel procedure for the identification of chaos in complex biological systems  [\onlinecite{Bazeia,Ramos3}]. However, this method has not been tested on any type of nanodevices from the point of view of the tight-binding model and experiments on nanowires.

In nanowires research, there is a significant interest in measuring the correlation length with high precision. From this length, one are able to extract the phase-coherence length ($l_\phi$) using the following relation to the quasi-one-dimension nanowire [\onlinecite{Lee,HoutenPRB,Houten}]
\begin{eqnarray}
l_\phi = \frac{\Phi_c}{\Gamma_\perp d},\label{lphi}
\end{eqnarray}
where $\Phi_c = 0.42 \times \Phi_0$. The $\Phi_0 = h/e$ and $d$ are the magnetic flux quanta and the nanowire diameter, respectively. However, the correlation width length ($\Gamma_\perp$) is usually obtained experimentally from the correlation function [\onlinecite{Doh,Alagha, Kim,Elm}].

In this work, we perform a complete analysis of the universal conductance fluctuations from the perspective of the tight-binding model of a quasi-one-dimensional nanowire submitted to a perpendicular and parallel magnetic field. Hence, the conductance peaks density model was tested comparing the results obtained by counting the number of maxima of the universal conductance fluctuation with the the results obtained by correlation function Eqs.(\ref{rhopp}) and (\ref{rhopr}). To conclude our study, we applied for the first time that methodology in the experimental magnetoconductance data of an InAs nanowire submitted to perpendicular and parallel magnetic field. The experimental data was obtained from measurements documented in Ref.[\onlinecite{Doh}].

\section{The Microscopic Model} 

We begin with the Hamiltonian model of a disordered quasi-one-dimensional semiconductor nanowire
\begin{eqnarray}
\mathcal{H}=\frac{1}{2m_{eff}}\bigg({\bf p}-e{\bf A}\bigg)^2 + \mu{\bf B}\cdot \mathbf{\sigma}+U({\bf r}),
\end{eqnarray}
where  ${\bf A}$ is the potential vector corresponding to the component of the magnetic field perpendicular to  the nanowire, ${\bf B}$ is the magnetic field, $\mu$ is the magnetic moment, $m_{eff}$ is the effective mass, $\mathbf{\sigma}=\left(\sigma_x,\sigma_y,\sigma_z\right)$ is the vector of Pauli matrix, and $U({\bf r})$ is the electrostatic disorder potential. 

\begin{figure}
\includegraphics[width=7cm,height=10cm]{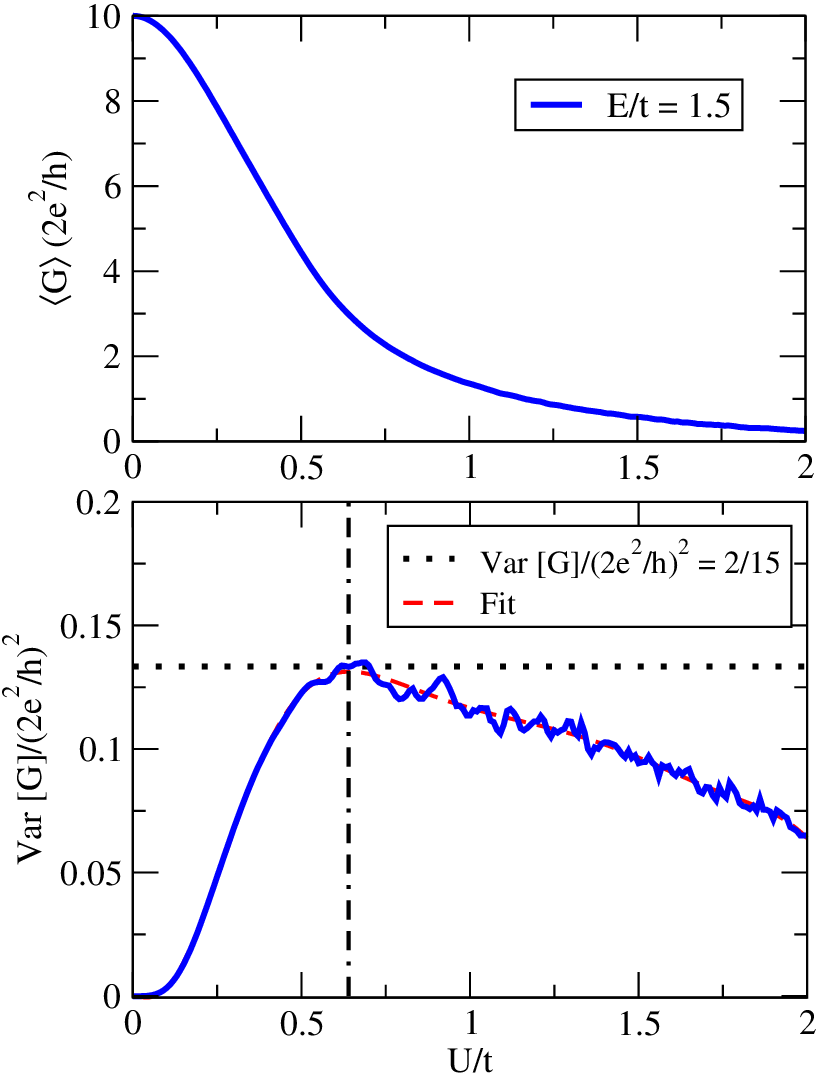} \quad \quad \quad
\caption{ The average (up) and variance (down) of conductance in function of electrostatic potential $U/t$ to fix Fermi energy $E = 1.5 t$. They were obtained from 15,000 disorder realizations. The numerical data of variance were fitted and found that the universal valuer of variance arises at $U=0.65 t$.  }\label{figura1}
\end{figure}

Our aim is to study the universal conductance fluctuation of a nanowire. Accordingly it is convenient to use the Landaur-B\"uttiker formulation at low temperature given by
\begin{eqnarray}
G=\frac{e^2}{h} \mathcal{T},
\end{eqnarray}
where $\mathcal{T}$ is the transmission coefficient between the reservoirs of charges connected to nanowire on the left (L) and right (R) sides. The  transmission coefficient is obtained from the scattering matrix 
\begin{eqnarray}
\mathcal{S}=
\left[\begin{array}{cc}
r&t' \\
t &  r'
\end{array}\right], \label{Smatrix}
\end{eqnarray}
as $\mathcal{T}=\mathbf{Tr}\left[t^\dagger t\right]$, where $t(t')$ and $r(r')$ are the transmission and the reflection matrix blocks, respectively. 
Furthermore, in the recursive Green's function framework, the transmission coefficient can be written as $\mathcal{T}=\mathbf{Tr}\left[\mathbf{\Gamma}_L\mathbf{G}^{r}_{LR}\mathbf{\Gamma}_R\mathbf{G}^{a}_{LR}\right]$, where $\mathbf{G}^{r,a}_{LR}$ are the retarded and advanced Green's functions which describe the disordered nanowire and $\mathbf{\Gamma}_{L,R}$ is the level width matrices that connect the device to the left or right reservoirs [\onlinecite{Akkermans,Caio,Qiao}]. In presence of magnetic field, the tight-binding Hamiltonian is written as a function of the creation and annihilation operators as in the following
\begin{eqnarray}
\mathcal{H}=-t \sum_{ij}  e^{i\theta_{ij}} c_i^\dagger c_j + \left(4t + \mu  {\bf B}\cdot \mathbf{\sigma}\right) \sum_{i} c_i^\dagger c_i,\label{TB}
\end{eqnarray}
where $t=\hbar^2/(2m_{ef}a^2)$ is the nearest neighbor hopping energy, $a$ is the square lattice constant, and $\theta_{ij}$ is obtained from vector potential as following $\theta_{ij}=-e/\hbar \int_i^j {\bf A}\cdot d{\bf l}$. 

\begin{figure}
\includegraphics[width=7cm,height=5cm]{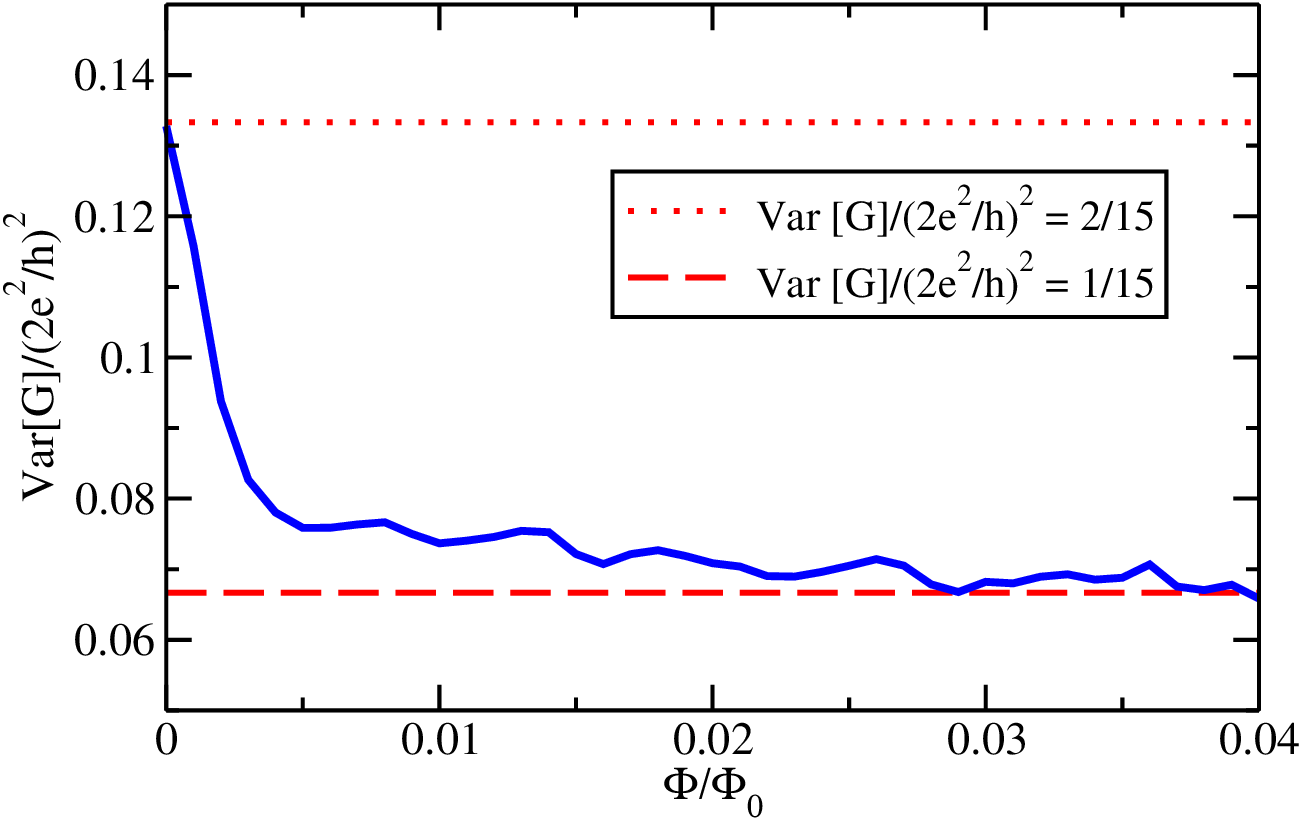} \quad \quad \quad
\caption{The variance of conductance in function of perpendicular magnetic flux to fix Fermi energy $E = 1.5 t$ and electrostatic potential $U=0.65t$ from 15,000 disorder realizations. As expected, the variance does a crossover between $2/15$ to $1/15$.}\label{figura11}
\end{figure}

The quasi-one-dimension nanowire has been numerically simulated by a square lattice with length $L=310 a$ and width $W=25 a$ in the $x$ and $y$ direction, respectively. Moreover, the disorder in the square lattice is realized by an electrostatic potential $U$ which varies randomly from site to site uniformly distributed in the interval $\left(-U/2,U/2\right)$. In order to develop our numeric calculation, the Kwant software has been used [\onlinecite{Kwant}]. 

\begin{figure*}[!]
\includegraphics[width=16cm,height=13cm]{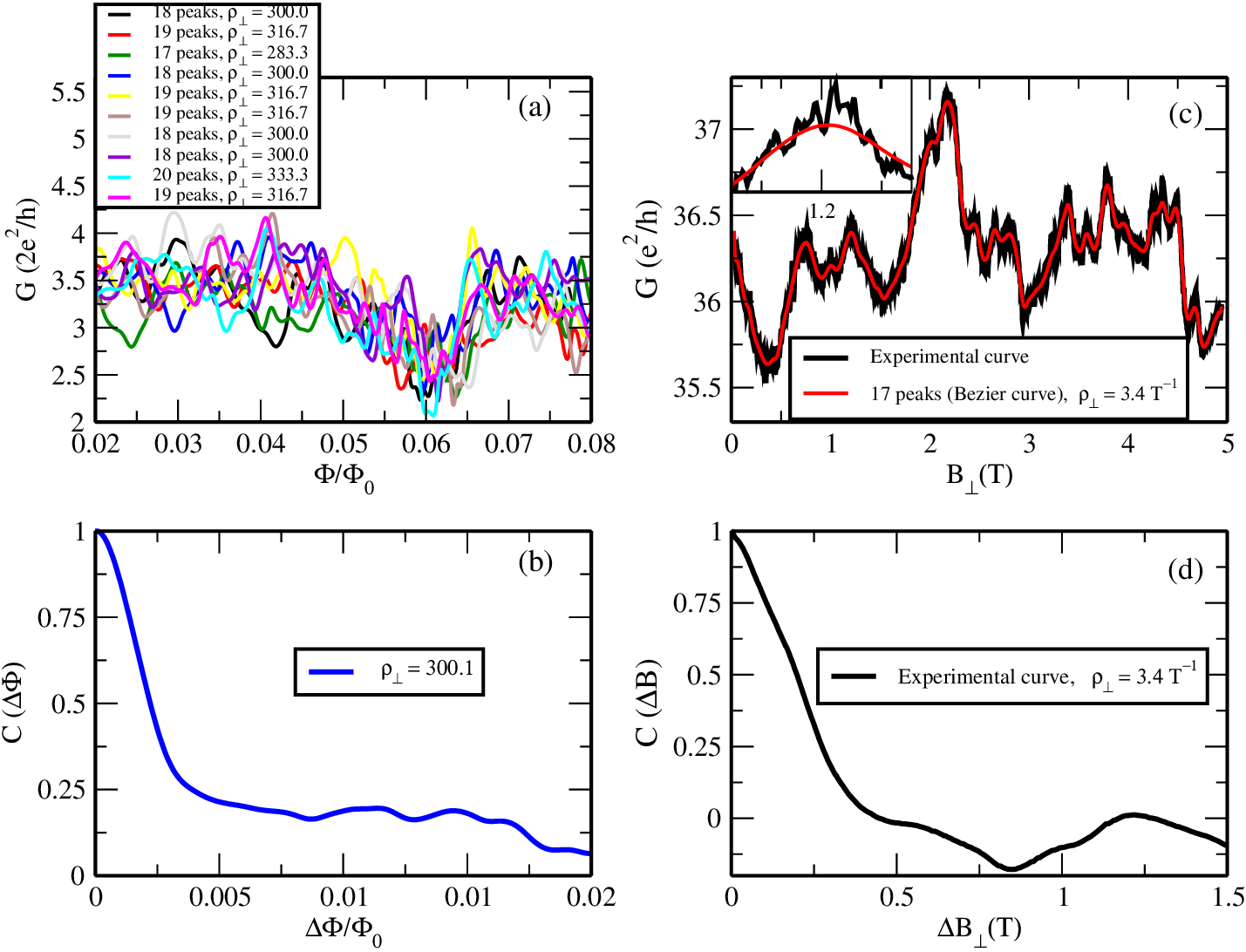} \quad \quad \quad
\caption{(a) Ten typical conductance curves in function of adimensional perpendicular magnetic flux, each one is for a single disorder realization with $U=0.65t$, $E=1.5t$ and magnetic flux steps of $5 \times 10^{-5}$. The parameter $\rho_\perp$ is the density of maxima (number of peak over range perpendicular magnetic flux). (b) Conductance correlation in function of perpendicular magnetic flux obtained from 1,000 realizations. (c) The black curve is a typical experimental magnetoconductance at $30$ mK of an InAs nanowire with length $L=107$ nm submitted to perpendicular magnetic field and the red curve is the B\'ezier fit. (d) Magnetoconductance correlation in function of perpendicular magnetic field obtained from experimental curve.}\label{figura2}
\end{figure*}

\begin{figure*}[!]
\includegraphics[width=16cm,height=13cm]{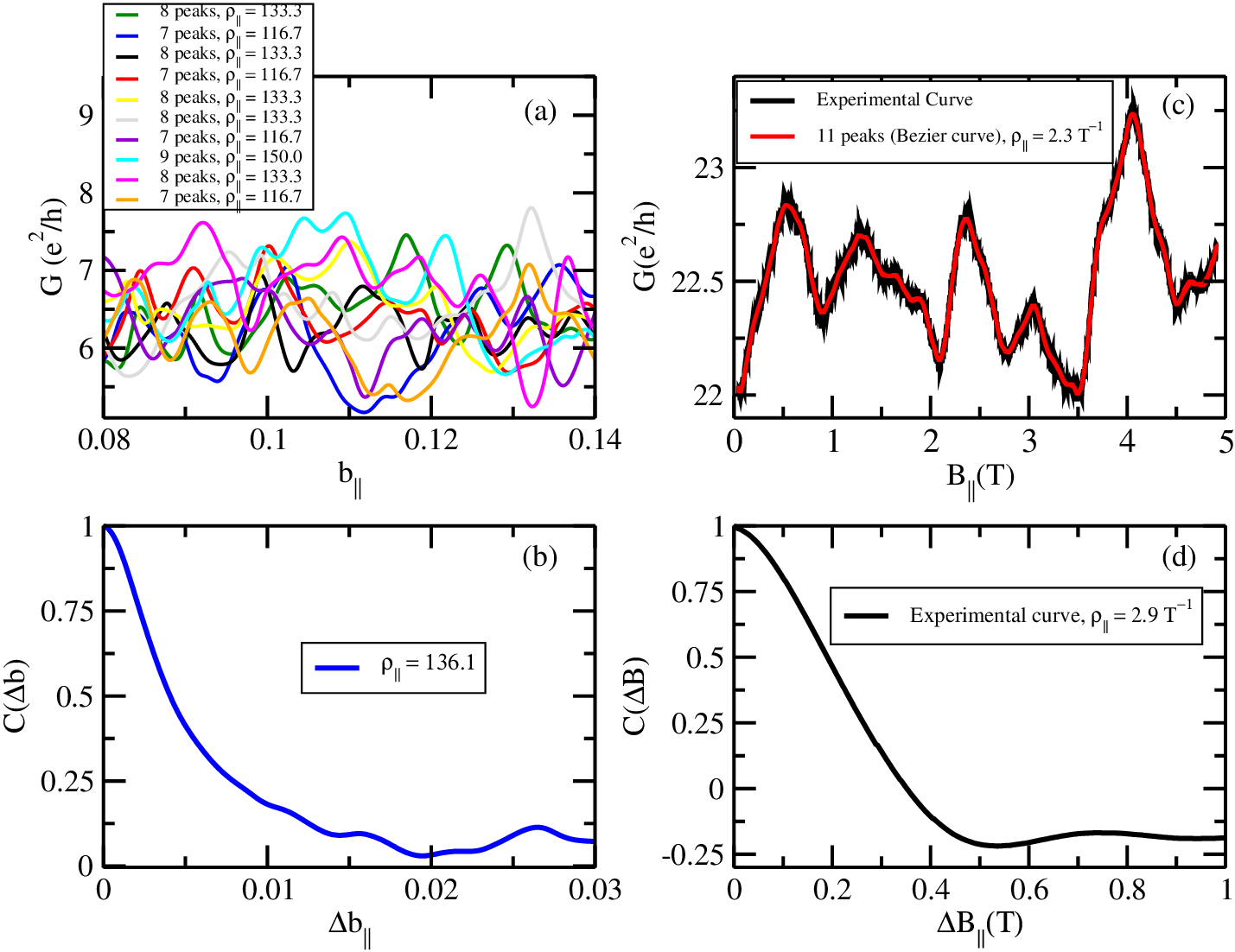} 
\caption{(a) Ten typical conductance curves in function of adimensional parallel magnetic flux, each one is for a single disorder realization with $U=0.65t$, $E=1.5t$ and magnetic flux steps of $5 \times 10^{-5}$. The parameter $\rho_{||}$ is the density of maxima (number of peak over range parallel magnetic flux). (b) Conductance correlation in function of parallel magnetic flux obtained from 1,000 realizations. (c) The black curve is a typical experimental magnetoconductance at $30$ mK of an InAs nanowire with length $L=440$ nm submitted to parallel magnetic field and the red curve is the B\'ezier fit. (d) Magnetoconductance correlation in function of parallel magnetic field obtained from experimental curve.}\label{figura3}
\end{figure*}

\section{Results} 

In the universal regime, the quasi-one-dimension nanowire has the variance of conductance given by $var[G]/(2e^2/h)^2 = 2/15$ in the absence of magnetic field, i.e. preserving the time-reversal symmetry [\onlinecite{Beenakker}]. Therefore, we develop a numerical calculation to find the best width electrostatic potential $U$ in units of $t$ which supports the universal regime. In the Fig.(\ref{figura1}) it is depicted the average and variance of conductance as a function of $U/t$ with fix Fermi energy $E=1.5 t$ and without magnetic field. They were obtained from 15,000 disorder realizations. Fitting the variance numerical data, we found that the universal variance value arises at $U=0.65 t$. After obtaining the best width electrostatic potential, we are enabled to analyze the conductance peak density as a function of perpendicular and parallel magnetic fields.

We begin by analyzing the universal conductance fluctuations as a function of the adimensional perpendicular magnetic flux $\Phi/\Phi_0=B_\perp a^2/(h/e)$, where we take  ${\bf B}=\left(0,0,B_\perp\right)$, ${\bf A} = \left(-B_\perp y,0,0\right)$ in the Eq.(\ref{TB}). In the Fig. (\ref{figura11}) was plotted the variance of conductance in function of perpendicular magnetic flux to $U=0.65t$ and $E=1.5t$ from 15,000 disorder realizations. The variance does a crossover between $2/15$ to $1/15$ because of the breaking of time-reversal symmetry, as expected [\onlinecite{Beenakker}]. Furthermore, the Fig.(\ref{figura2}.a) shows ten typical curves of conductance, each one for a single disorder realization with $U=0.65t$, $E=1.5t$ {\it and magnetic flux steps of $5 \times 10^{-5}$}. From these typical curves, we were able to count the number of maxima. Dividing the number of maxima by the range of perpendicular magnetic flux ($\Delta(\Phi/\Phi_0)=0.06$) one obtains the average of density peak $\left\langle \rho_\perp \right\rangle = 308.3 $.

We can also calculate the peak density using the Eq.(\ref{rhopp}). However, firstly it is necessary to obtain the correlation length, $\Gamma_\perp$, defined as the half height of correlation function. The procedure requires a large number of realizations to be evaluated. The Fig.(\ref{figura2}.b) shows the correlation function obtained from 1,000 disorder realizations as a function of the perpendicular magnetic flux. From the Fig.(\ref{figura2}.b), one found a correlation length $\Gamma_\perp = 2.25 \times 10^{-3}$. Replacing in Eq.(\ref{rhopp}) the peak density is $\left\langle \rho_\perp \right\rangle = 300.1$ which is in accordance to the result obtained counting the number of maxima.

After the analyze of the conductance peaks density in a quasi-one-dimension nanowire using a numerical point of view, we apply, for the first time, the methodology to experimental data of a semiconductor nanowire. The Fig.(\ref{figura2}.c) shows a typical experimental magnetoconductance data at $30$ mK of a InAs nanowire with length $L=107$ nm submitted to perpendicular magnetic field. 

Although the typical numeric curves of Fig.(\ref{figura2}.a) and experimental curve of Fig.(\ref{figura2}.c) have an apparently similar behavior, one can not count the number of peaks directly of the latter. The inset display of Fig.(\ref{figura2}.c) shows that the experimental curve has a random noise background which is not present in the Fig.(\ref{figura2}.a). The background appears in the experimental magnetoconductance data due to the thermal noise and the experimental apparatus. The random background is irrelevant to the process under investigation, being necessary a smoothing in its behavior. 

To smooth the experimental conductance data of Fig.(\ref{figura2}.c), we apply the B\'ezier algorithm [\onlinecite{bezier}] and the result is displayed in red color, the conductance as a function of the perpendicular magnetic field. To obtain the red curve in the Fig.(\ref{figura2}.c), we developed the Bezier algorithm on experimental data the number of times necessary to the peak number converge. The inset display of Fig.(\ref{figura2}.c) shows that the B\'ezier curve eliminates the random experimental data background holding only the general chaotic behavior. The procedure ensures the count of the number of maxima from the B\'ezier fit obtained from the average of peak density $\left\langle \rho_\perp \right\rangle = 3.4$ T$^{-1}$, where the range of perpendicular magnetic field is $\Delta B_\perp = 4.915$ T. 

After to obtain the  average of peak density from B\'ezier fit, we are able to estimate the phase-coherence length of InAs nanowire with $d=80$ nm [\onlinecite{Doh}]. Replacing $\left\langle \rho_\perp \right\rangle = 3.4$ T$^{-1}$ in Eq.(\ref{rhopp}), we obtain that $\Gamma_\perp=0.1952$ T. Hence, we can estimate from Eq. (\ref{lphi}) that the phase-coherence length is $l_\phi \approx 111$ nm.

In order to confirm the result obtained by counting the number of maxima, the correlation function was calculated from the experimental data showed in Fig. (\ref{figura2}.d). Using the latter, we obtain a correlation length of $\Gamma_\perp=0.1975$ T. Replacing in Eq.(\ref{rhopp}) one obtain the peak density $\left\langle \rho_\perp \right\rangle = 3.4$ T$^{-1}$, which is in nice accordance with the result obtained by counting the number of maxima. 

As a second preponderant result, we analyze the conductance peak density as a function of the adimensional parallel magnetic field $b_{||} =\mu B_{||}/t$, where it is taken ${\bf B}=\left(B_{||},0,0\right)$ in the Eq.(\ref{TB}). The Fig.(\ref{figura3}.a) shows ten typical curves of conductance, each one for a single disorder realization with $U=0.65t$, $E=1.5t$ {\it and magnetic flux steps of $5 \times 10^{-5}$}. We can realize that peak density of Fig.(\ref{figura3}.a) is smaller than the one of the Fig.(\ref{figura2}.a), which is according to theoretical results Eqs.(\ref{rhopp}) and (\ref{rhopr}), and hence with the experimental data of nanowire Figs.(\ref{figura2}.c) and (\ref{figura3}.c). 

From the typical curves of Fig.(\ref{figura3}.a), we count the number of maxima. Dividing the number of maxima by the range of parallel magnetic flux ($\Delta(b_{||})=0.06$)  one obtains the average of peak density $\left\langle \rho_{||} \right\rangle = 128.3$. Moreover, the Fig.(\ref{figura3}.b) shows the correlation function  obtained from 1,000 disorder realizations in function of parallel magnetic flux. From the latter, it was found that the length correlation is $\Gamma_{||} = 4.05  \times 10^{-3}$. Replacing in the Eq.(\ref{rhopr}), one shows that the peak density is  $\left\langle \rho_{||} \right\rangle = 136.1$ which is in nice accordance with the result obtained by counting the number of maxima.

In the Fig.(\ref{figura3}.c), a typical experimental magnetoconductance data at $30$ mK of a InAs nanowire with length $L=440$ nm submitted a parallel magnetic field was plotted. As discussed previously, we cannot count the number of maxima from the experimental data because of intrinsic noise white noise background. Hence, we use again the B\'ezier algorithm to display in red the conductance as a function of the parallel magnetic field, as it can be seen in Fig.(\ref{figura3}.c). From the latter, we were able to count the number of maxima and obtain the average of peak density $\left\langle \rho_{||} \right\rangle = 2.3$ T$^{-1}$, where the range of parallel magnetic field is $\Delta B_{||} = 4.859$ T. Furthermore, the Fig.(\ref{figura3}.d) shows the correlation function obtained from experimental data of Fig.(\ref{figura3}.c). From the latter, we found the correlation length $\Gamma_{||} = 0.1901$ T. Replacing this result in the Eq.(\ref{rhopr}) one holds that the peak density is $\left\langle \rho_{||} \right\rangle = 2.9$ T$^{-1}$ which is in accordance to the result obtained by counting the number of maxima.

\section{Conclusion}

In summary, we have done a complete analysis of universal conductance fluctuation in quasi-one-dimensional nanowire submitted to perpendicular and parallel magnetic fields using the tight-binding model. We use the conductance peak density model showing a satisfactory accordance between the model, Eq.(\ref{rhopp}) and (\ref{rhopr}), and numeric calculation. Furthermore, we applied successfully for the first time that the peak density model in the experimental magnetoconductance data of as InAs nanowire submitted to perpendicular and parallel magnetic field at $30$ mK. 

We conclude that the method is an efficient alternative to obtain the correlation width length, which is normally obtained experimentally from correlation function [\onlinecite{Doh,Alagha, Kim,Elm}]. We believe that our work can support a myriad of future works getting more precise results for correlation length and, consequently, the  phase-coherence length in nanowires. As we have shown, the application of the theory can be fundamental in the obtaining and characterization of the universal regime, of the coherence length responsible for the various quantum phenomena including entanglement. Therefore, this method and their respective developments deserves attention in spintronics, graphene flakes and topological isolator where  phase-coherence length is a relevant experimental parameter [\onlinecite{Ramos2,Lundeberg,Horsell,Terasawa,Choe,Wang,Vasconcelos,Vasconcelos2,Ramos3,Lima,Hagymasi,Gopar,Diez}].

This work was partially supported by CNPq, CAPES and FACEPE (Brazilian Agencies).

\end{document}